\documentclass[11pt,letterpaper]{article}
\usepackage[T1]{fontenc}
\usepackage{lineno}
\usepackage{amsfonts,amsmath,amsthm,pifont,amssymb,dsfont,mathtools,algpseudocode}
\usepackage[margin=1in]{geometry}
\usepackage[colorlinks,citecolor=blue,linkcolor=blue,urlcolor=red]{hyperref}
\usepackage{url}
\usepackage[linesnumbered,boxed,ruled,vlined]{algorithm2e}
\usepackage{thm-restate}
\usepackage{enumitem}
\usepackage{etoolbox}
\usepackage{graphicx}
\usepackage{mathdots}
\usepackage[normalem]{ulem}
\usepackage{xspace}
\usepackage{authblk}
\xspaceaddexceptions{]\}}
\usepackage{comment}
\usepackage{tabu}
\usepackage{framed}
\usepackage{float,wrapfig}
\usepackage{subfigure}
\usepackage{tikz}
\usepackage[draft]{fixme}
\usepackage[capitalise]{cleveref}

\newcommand*\linenomathpatch[1]{%
  \cspreto{#1}{\linenomath}%
  \cspreto{#1*}{\linenomath}%
  \csappto{end#1}{\endlinenomath}%
  \csappto{end#1*}{\endlinenomath}%
}

\linenomathpatch{equation}
\linenomathpatch{gather}
\linenomathpatch{multline}
\linenomathpatch{align}
\linenomathpatch{alignat}
\linenomathpatch{flalign}

\usetikzlibrary{arrows,arrows.meta,decorations.pathmorphing}

\theoremstyle{plain}
\newtheorem{theorem}{Theorem}[section]
\newtheorem{lemma}[theorem]{Lemma}

\theoremstyle{definition}

\fxsetup{mode=multiuser,theme=color, layout=inline}
\FXRegisterAuthor{tk}{atk}{\color{brown} {\bf Tomasz}}

\setlist[enumerate]{nosep, topsep=1ex}
\setlist[itemize]{nosep, topsep=1ex}
\setlist[description]{nosep}
\allowdisplaybreaks
\def\ShowAuthNotes{1}
\ifnum\ShowAuthNotes=1
\newcommand{\authnote}[2]{\ \\ \textcolor{red}{\parbox{0.9\linewidth}{[{\footnotesize {\bf #1:} { {#2}}}]}}\newline}
\else
\newcommand{\authnote}[2]{}
\fi

\newcommand{\dd}{\mathinner{.\,.\allowbreak}}

\DeclareMathOperator{\SA}{SA}

\DeclareMathOperator{\LCE}{LCE}
\DeclareMathOperator{\ISA}{ISA}

\title{Tiny Suffix Arrays
}

\author{
Sharma V. Thankachan
}

\affil{
Department of Computer Science \\ North Carolina State University, Raleigh, NC, USA \\{svalliy@ncsu.edu}
}

\begin{document}
	\maketitle
	\thispagestyle{empty}

\begin{abstract}
Suffix array and inverse suffix array queries on a text of
length $n$ can be answered in $\widetilde{O}(\tau^3)$ time
using an auxiliary data structure occupying $O(n/\tau)$
words, for any $\tau \in [1,n]$.
\end{abstract}

\section{Introduction}\label{sec:intro}
Let $T[0 \dd n)$ be a text over an ordered alphabet of size
$\sigma \leq n$. Without loss of generality, we assume that
the alphabet is $\{0,1,\ldots,\sigma-1\}$ and that $T$ is
terminated by a unique sentinel character $\$$ (corresponding to $0$), which is
lexicographically smaller than all other characters.
The \emph{suffix array} ($\SA$) of $T$ is a permutation of $[0 \dd n)$ that lists the starting positions of all suffixes of $T$ in lexicographic order. The \emph{inverse suffix array} ($\ISA$) is the inverse permutation of $\SA$, i.e., $\SA[i]=j$ if and only if $\ISA[j]=i$. For example, for $T[0 \dd 7)=\texttt{banana\$}$, we have $\SA=[6,5,3,1,0,4,2]$ and $\ISA=[4,3,6,2,5,1,0]$. These fundamental data structures form the basis of numerous algorithms in string processing, data compression, bioinformatics, and full-text indexing~\cite{gusfield1997algorithms,DBLP:journals/siamcomp/ManberM93,DBLP:conf/focs/Weiner73}.

Storing $\SA$ and $\ISA$ explicitly requires $\Theta(n\log n)$ bits each, whereas the text itself can be represented in only $n\lceil\log\sigma\rceil$ bits. This difference is particularly significant for small alphabets (e.g., DNA sequences), where the space occupied by these  structures can be substantially larger than that of the text. A major breakthrough in addressing this problem came with the introduction of the \emph{Compressed Suffix Array} (CSA)~\cite{DBLP:journals/siamcomp/GrossiV05} and the \emph{FM-index}~\cite{DBLP:journals/jacm/FerraginaM05}, which showed that $\SA$ and $\ISA$ queries can be supported in $\tilde{O}(1)$ time using only $O(n\log\sigma)$ bits, where $\tilde{O}$ hides polylogarithmic factors in $n$.
These foundational results initiated an extensive line of research, leading to several breakthroughs, including compressed suffix trees, entropy-compressed representations, and repetitiveness-aware representations~\cite{DBLP:journals/talg/BelazzouguiN14,DBLP:conf/soda/FerraginaM04,DBLP:conf/soda/GrossiGV03,DBLP:journals/jacm/GagieNP20,DBLP:conf/focs/KempaK23,DBLP:conf/soda/KempaK26,DBLP:journals/csur/Navarro21,kempa2026constantISA,DBLP:journals/mst/Sadakane07,russo2008fully} of these fundamental data structures. 
Despite these remarkable advances, all such representations require $\Omega(n\log\sigma)$ bits in the worst case. In fact, this is an information-theoretic lower bound for
representing either $\SA$ or $\ISA$, since the  text can be reconstructed from either permutation together with the frequency of each character. Thus, the space occupied by the text itself represents a fundamental
barrier to space-efficient encodings of the $\SA$ and $\ISA$.
When the text is available, however, we can focus
exclusively on the auxiliary space required to encode these arrays.
This leads to the following natural problem:
\begin{center}
  \emph{Design a data structure within a prescribed space budget that efficiently supports $\SA$ and $\ISA$ queries, given read-only access to the underlying text.}  
\end{center}

Several related problems have been studied under a bounded auxiliary-space budget of $O(n/\tau)$ words, for a tradeoff parameter $\tau \geq 1$. In particular, Bille et al.~\cite{bille2014time} showed that longest common extension (LCE) queries can be supported deterministically in $O(\tau^2)$ time using $O(n/\tau)$ words of auxiliary space. Subsequently, Bille et al.~\cite{BilleEtAlLCE2015} improved the query time to $O(\tau)$ while retaining the same space bound. Further results on LCE queries and their efficient construction were obtained by Tanimura et al.~\cite{DBLP:conf/cpm/TanimuraIBIPT16}, Birenzwige et al.~\cite{DBLP:journals/siamcomp/BirenzwigeGP25}, and Kosolobov and Sivukhin~\cite{DBLP:conf/cpm/KosolobovS24}. Related results on sparse suffix-tree construction under small space constraints were obtained by Gawrychowski and Kociumaka~\cite{DBLP:conf/soda/GawrychowskiK17} and Kosolobov and Sivukhin~\cite{DBLP:conf/cpm/KosolobovS24}.
Meanwhile, significant progress has been made in text indexing under small auxiliary-space budgets. Bathie et al.~\cite{DBLP:conf/cpm/BathieCS24} developed a data structure for internal pattern matching using $\tilde{O}(n/\tau)$ words of auxiliary space, supporting queries in $O(\tau+\log n\log^3\log n)$ time. In a breakthrough result, Bibbens, Borevitz, and McCauley~\cite{BibbensBorevitzMcCauleySTOC26} developed a text index based on function inversion~\cite{DBLP:journals/siamcomp/FiatN99} that occupies $O(n/\tau)$ words of auxiliary space and reports all $occ$ occurrences of a pattern of length $m$ in $\tilde{O}(m+\tau^4+\tau^3\cdot occ)$ time. Subsequently, Charalampopoulos et al.~\cite{CharalampopoulosEtAl2027} improved the query time to $\tilde{O}(m+\tau^3/m^2+(1+\tau^2/m)\cdot occ)$ while retaining the same space bound.\footnote{The original trade-off in~\cite{CharalampopoulosEtAl2027}
is $\tilde{O}(\tau)$ words of space and
$\tilde{O}(m+(n/\tau)^3/m^2+(1+(n/\tau)^2/m)\cdot occ)$ query time
for any $\tau \in [1,n]$.
The bound we stated follows by substituting $\tau\log^c n$ for $n/\tau$,
for a suitable constant $c$.}

A step toward supporting SA and ISA queries in small auxiliary space was taken in recent work by Thankachan~\cite{ThankachanFOCS26}, where it was shown that SA, ISA, and several fundamental suffix-tree operations can be supported in polylogarithmic time using $O(n\log\log\log\sigma)$ bits of auxiliary space, where $\sigma$ is the alphabet size.
 However, this bound remains linear in $n$ up to an alphabet-dependent factor. To the best of our knowledge, no prior work establishes a general time-space tradeoff for supporting both SA and ISA queries using $O(n/\tau)$ words of auxiliary space for an arbitrary parameter $\tau \geq 1$. We obtain the following result.


\begin{theorem}\label{thm:main}
Let $T[0 \dd n)$ be a text over an ordered alphabet of size $\sigma \leq n$, accessible through a read-only oracle supporting character access in constant time. For any integer $1 \leq \tau \leq n$, there exists an auxiliary data structure occupying $O(n/\tau)$ words that supports both suffix array $(\SA)$ and inverse suffix array $(\ISA)$ queries in $\tilde{O}(\tau^3)$ time.
\end{theorem}


\vspace{-6mm}
\section{Building Blocks}
\vspace{-2mm}

We begin with a few useful lemmas. Throughout this section, we assume that the text $T$ is accessible through a read-only oracle supporting character access in ${O}(1)$ time. We also maintain the auxiliary data structure of Bille et al.~\cite{BilleEtAlLCE2015}, which occupies $O(n/\tau)$ space and supports longest common extension (LCE) queries in ${O}(\tau)$ time. Specifically, an LCE query $\LCE(i,j)$ returns the length of the longest common prefix of the suffixes $T[i\dd n)$ and $T[j\dd n)$.

\begin{lemma}\label{lem:longLCE}
There exists an auxiliary data structure occupying $O(n/\tau)$ space that, given two positions $i,j \in [0 \dd n)$ satisfying $\LCE(i,j) \geq \tau$, reports all $k$ suffixes of $T$ lying strictly between $T[i \dd n)$ and $T[j \dd n)$ in lexicographic order in $\tilde{O}(\tau^2+k)$ time.
\end{lemma}

\begin{proof}
We adapt existing techniques based on suffix sampling and  range queries~\cite{DBLP:journals/vldb/AyadLP25,DBLP:journals/algorithmica/ChienHSTV15}. First, we sample the suffixes $T[a \dd n)$ and the corresponding prefixes $T[0 \dd a)$ at positions $a=\tau,2\tau,3\tau,\ldots$. For each sampled position $a$, we construct a tuple $(a,x_a,y_a)$, where $x_a$ is the lexicographic rank of $T[a \dd n)$ among the sampled suffixes, and $y_a$ is the co-lexicographic rank of $T[0 \dd a)$ among the sampled prefixes. 
We maintain a two-dimensional orthogonal range reporting structure on the points $(x_a,y_a)$, storing $a$ as satellite information. We use the linear-space data structure of Chazelle~\cite{DBLP:journals/siamcomp/Chazelle88}, which, for a set of $N$ points, occupies $O(N)$ words and reports $k$ points in $O(\log N+k\log^\varepsilon N)$ time, for any fixed constant $\varepsilon>0$. The total space usage is therefore $O(n/\tau)$ words.

Given a query $(i,j)$, assume without loss of generality that $T[i \dd n)<T[j \dd n)$, and let $s=\LCE(i,j)\geq\tau$. For each $d\in[0\dd\tau)$, we issue a two-dimensional range reporting query to identify sampled positions $a$ satisfying (i) $T[a-d\dd a)=T[i\dd i+d)$, and (ii) $T[i+d\dd n)<T[a\dd n)<T[j+d\dd n)$. Since $d\leq s$, these conditions are equivalent to $T[i\dd n)<T[a-d\dd n)<T[j\dd n)$. The first condition defines a contiguous interval of co-lexicographic ranks, which can be obtained by binary searching for the reversed pattern $T[i\dd i+d)$ among the reversed sampled prefixes. Each comparison takes $O(d)$ character accesses. Similarly, the second condition defines a contiguous interval of lexicographic ranks, obtained by binary searching for $T[i+d\dd n)$ and $T[j+d\dd n)$ among the sampled suffixes. Here, instead of comparing suffixes character by character, we use LCE queries to determine their lexicographic order in $\tilde{O}(\tau)$ time per comparison. Thus, both intervals can be computed in $\tilde{O}(\tau)$ time. We report the points in their Cartesian product, outputting $a-d$ for each reported point.
Each range reporting query takes $\tilde{O}(\tau+k_d)$ time, where $k_d$ is its output size. Position $0$ and the at most $\tau$ positions following the last sample are handled directly in $\tilde{O}(\tau^2)$ time. Since every other position is reported for exactly one value of $d$, the overall  time is $\tilde{O}(\tau^2+k)$. 
\end{proof}

\begin{lemma}\label{lem:PM}
There exists an auxiliary data structure occupying $O(n/\tau)$ space that, given a pattern $P[0 \dd m)$ and a character range $[\alpha,\beta]$, reports all occurrences of $Pc$ (i.e., $P$ appended with  $c$), for every $c\in[\alpha,\beta]$, in $\tilde{O}(m+\tau^3/m^2+(1+\tau^2/m)\cdot occ)$ time, where $occ$ is the output size.
\end{lemma}

\begin{proof}
A straightforward approach is to maintain the pattern matching data structure of Charalampopoulos et al.~\cite{CharalampopoulosEtAl2027}, occupying $O(n/\tau)$ space, and issue a separate pattern reporting query for $Pc$, for each $c\in[\alpha,\beta]$. However, this results in a total query time of $O(\sigma(m+\tau^3/m^2)+(1+\tau^2/m)occ)$. To achieve the claimed bound, we instead view $T$ as a string over the alphabet $\{\#,0,1\}$, replacing each character $c$ with $\#$ followed by its binary representation of length $\lceil\log\sigma\rceil$. Let $T'$ denote the resulting string, of length $n'=O(n\log\sigma)$. We do not store $T'$ explicitly, since any character of $T'$ can be recovered from $T$ in $\tilde{O}(1)$ time. We maintain the data structure of Charalampopoulos et al.~\cite{CharalampopoulosEtAl2027} on $T'$, using the parameter $\tau'=\Theta(\tau\log\sigma)$, so that the auxiliary space remains $O(n'/\tau')=O(n/\tau)$.

Given a query $(P,[\alpha,\beta])$, let $P'$ be the encoding of $P$ under the same transformation. We partition $[\alpha,\beta]$ into $O(\log\sigma)$ disjoint intervals, each consisting of characters whose binary representations share a common prefix $B$. For each such interval, we issue a pattern reporting query for $P'\#B$ on $T'$. Since these intervals partition $[\alpha,\beta]$, the reported occurrences correspond precisely to those of $Pc$ for all $c\in[\alpha,\beta]$, without duplication. The number of pattern reporting queries is thus reduced from $O(\sigma)$ to $O(\log\sigma)$, while the transformed pattern lengths are $O(m\log\sigma)$. Substituting these parameters into the query bound yields the claimed time complexity. 
\end{proof}

\vspace{-3mm}
\section{The Data Structure }
Depending on the value of $\tau$, we consider two cases.
\subsection{Solution for $\tau>\sqrt{n}$}
In this case, our data structure consists only of
the $O(n/\tau)$-word structure supporting $\LCE$
queries, together with at most $O(n/\tau)$ words
of additional working space.

To answer an $\SA[i]$ query, we need to identify
the $i$-th lexicographically smallest suffix, for
which we use known selection algorithms.
Since the classical median-of-medians algorithm
requires $O(n)$ working space, we instead employ
space-efficient selection algorithms as follows.
We consider two subcases depending on the value of $\tau$.

\begin{itemize}
\item 
For $\sqrt{n}<\tau\leq n/\log n$, we use
Frederickson's read-only selection algorithm~\cite{DBLP:journals/jcss/Frederickson87},
which requires $O(n\log n)$ comparisons and
$O(\log n)$ words of working space.
This fits within our $O(n/\tau)$-word space bound.
Since each comparison can be performed using an
$\LCE$ query in $O(\tau)$ time, the total query
time is $O(n\tau\log n) \subseteq\tilde{O}(\tau^3)$.

\item
For $\tau>n/\log n$, we instead use the
constant-workspace selection algorithm of
Munro and Raman~\cite{DBLP:journals/tcs/MunroR96}, which performs
$O(n^{1+\epsilon})$ comparisons for any fixed
constant $\epsilon>0$. Choosing $\epsilon=1/3$,
the resulting query time is $O(n^{4/3}\tau) \subseteq O(\tau^3)$.
\end{itemize}

To answer an $\ISA[j]$ query, we compare $T[j\dd n)$ with all other suffixes and count those that are lexicographically smaller. This requires $O(n)$
$\LCE$ queries and $O(1)$ working space, resulting
in a query time of $O(n\tau) \subseteq O(\tau^3)$,
since $\tau>\sqrt{n}$.

\vspace{-2mm}
\subsection{Solution for $\tau \leq \sqrt{n}$}

We first describe a scheme for sampling leaves of the suffix tree of $T$~\cite{DBLP:conf/focs/Weiner73}. Let $\ell_0,\ldots,\ell_{n-1}$ denote the leaves in left-to-right order, where $\ell_q$ corresponds to the suffix $T[\SA[q] \dd n)$. We initially sample the leaves $\ell_0,\ell_{\tau},\ell_{2\tau},\ldots$, together with $\ell_{n-1}$. For each initially sampled leaf whose suffix has length at least $\tau$, we consider the subtree containing exactly those suffixes sharing its first $\tau$ characters, and additionally sample the leftmost and rightmost leaves of this subtree. No additional sampling is needed for suffixes shorter than $\tau$. Clearly, the total number of sampled leaves is $O(n/\tau)$, and there are at most $\tau-1$ unsampled leaves between any two consecutive sampled leaves. The following lemma establishes a crucial property of this sampling scheme.

\begin{lemma}\label{lem:sampling}
If the suffixes corresponding to two consecutive sampled leaves have a longest common prefix (LCP) of length less than $\tau$, then every unsampled suffix between them has a LCP of length less than $\tau$ with each of these two suffixes.
\end{lemma}
\begin{proof}
Suppose an intermediate suffix shares a prefix of length at least $\tau$ with either endpoint. Since the endpoints have a LCP of length less than $\tau$, the corresponding length-$\tau$ prefix interval must have a boundary strictly between them. This boundary leaf would have been sampled, contradicting their consecutiveness.
\end{proof}

\vspace{-2mm}

Our auxiliary data structure consists of the following components: the structures of Lemmas~\ref{lem:longLCE} and~\ref{lem:PM}, the $O(n/\tau)$-word data structure of Bille et al.~\cite{BilleEtAlLCE2015} supporting $\LCE$ queries in $O(\tau)$ time, and a sampled suffix array (SSA). The SSA stores a list of pairs $(q,\SA[q])$, sorted by $q$, one for each sampled leaf $\ell_q$ selected by the sampling scheme described above. Since the number of sampled leaves is $O(n/\tau)$, the SSA occupies $O(n/\tau)$ words. Additionally, we reserve $O(\tau)$ words of working space, bringing
the total space to $O(n/\tau+\tau)$ words. This remains $O(n/\tau)$
since $\tau\leq\sqrt{n}$. 


\vspace{-2mm}

\subsubsection{Decoding $\SA[i]$}
Given a query index $i$, we binary search the SSA to check whether $\ell_i$ is sampled. If so, we return the stored value $\SA[i]$. Otherwise, we identify two consecutive sampled leaves $\ell_p$ and $\ell_q$ such that $p<i<q$, and retrieve $\SA[p]$ and $\SA[q]$ from the SSA. This takes $O(\log n)$ time. We now show how to decode the subarray $\SA[p+1\dd q-1]$, from which $\SA[i]$ can be directly retrieved. To this end, we compute $s=\LCE(\SA[p],\SA[q])$ in $O(\tau)$ time and consider the following two cases.

\begin{enumerate}
    \item 
We first consider the case where $s\geq\tau$.
We invoke Lemma~\ref{lem:longLCE} to report and store all suffixes lying strictly between $T[\SA[p]\dd n)$ and $T[\SA[q]\dd n)$. These correspond precisely to the entries of $\SA[p+1\dd q-1]$, possibly in a different order. Since there are $O(\tau)$ such suffixes, the reporting takes $\tilde{O}(\tau^2)$ time and $O(\tau)$ words of working space. We then sort the reported suffixes using $O(\tau\log\tau)$ comparisons, each performed via an LCE query followed by a character comparison in $\tilde{O}(\tau)$ time. Thus, the entire subarray can be decoded in $\tilde{O}(\tau^2)$ time.

\item 
We next consider the case where $s<\tau$.
Let $A=T[\SA[p]\dd n)$ and $B=T[\SA[q]\dd n)$.
By Lemma~\ref{lem:sampling}, every suffix strictly between
$A$ and $B$ has a longest common prefix with each endpoint
of length in $[s,\tau)$. We exploit this property to identify these suffixes using $O(\tau)$ pattern reporting queries.
Specifically, for each $r\in[s, \tau)$, we consider the prefix $A[0\dd r)$ and issue a query using Lemma~\ref{lem:PM} for all patterns $A[0\dd r)c$ with $c>A[r]$. Similarly, we consider $B[0\dd r)$ and query all patterns $B[0\dd r)c$ with $c<B[r]$. We restrict the queries at $r=s$ to characters strictly between $A[s]$ and $B[s]$, avoiding overlaps. These queries report precisely the suffixes lying strictly between $A$ and $B$, since every such suffix must differ from both endpoints within the first $\tau$ characters.

There are $O(\tau)$ queries, with pattern lengths at most $\tau$. By Lemma~\ref{lem:PM}, their total running time for these queries is
$\tilde{O}\left(\sum_{m=1}^{\tau}(m+\tau^3/m^2)+\tau^2 occ\right)
=\tilde{O}(\tau^3)$, since $occ=O(\tau)$.
We store the $O(\tau)$ reported suffix positions using $O(\tau)$ words of working space and
 sort them using $O(\tau\log\tau)$ LCE-based comparisons, taking $\tilde{O}(\tau^2)$ additional time. Thus, the subarray $\SA[p+1\dd q-1]$, and hence $\SA[i]$, can be obtained in $\tilde{O}(\tau^3)$ time.
\end{enumerate}
Combining both cases, we obtain a query time of $\tilde{O}(\tau^3)$. 

\subsubsection{Decoding $\ISA[j]$}

Given a query position $j$, we binary search the SSA, using LCE queries to compare $T[j\dd n)$ with the sampled suffixes and determine its lexicographic position among them. If $T[j\dd n)$ is itself sampled, we directly return its stored rank $\ISA[j]$. Otherwise, we identify two consecutive sampled leaves $\ell_p$ and $\ell_q$ such that $T[\SA[p]\dd n)<T[j\dd n)<T[\SA[q]\dd n)$ lexicographically. This takes $\tilde{O}(\tau\log n)$ time. We then decode the subarray $\SA[p+1\dd q-1]$ using the algorithm described in the previous subsection and locate $j$ within it to obtain $\ISA[j]$. Thus, ISA queries take $\tilde(\tau^3)$ time.

\section{Concluding Remarks}
We establish space--time trade-offs for suffix array and inverse
suffix array queries under a bounded auxiliary-space budget.
Together with existing results on LCE queries~\cite{BilleEtAlLCE2015}, this shows that
three of the most fundamental primitives in string algorithms
can be supported using sublinear auxiliary space.
Since suffix arrays form the basis of numerous string data
structures and algorithms (beyond exact pattern matching),
our result opens up opportunities for exploring new space--time
trade-offs across a broad range of string processing problems.




\paragraph{Acknowledgments.}
This research was supported in part by U.S. National Science Foundation (NSF)
Awards CCF-2316691 and CCF-2315822.
Many thanks to the authors of~\cite{CharalampopoulosEtAl2027}
for sharing their preprint and to Solon Pissis for helpful suggestions.

\paragraph*{Use of Generative AI} 
All original ideas, including the main results and proof techniques, were developed by the authors. OpenAI's ChatGPT was used to identify relevant literature, improve exposition, and review technical arguments. The authors take full responsibility for the correctness and content of the paper.

\bibliographystyle{abbrv}
\bibliography{main}

\end{document}